\documentclass{aastex}
\usepackage{spr-astr-addons}
\usepackage{subfigure}
\usepackage{url}

\RequirePackage{color}

\begin{document}

\title{SOUSA: the Swift Optical/Ultraviolet Supernova Archive}
\shorttitle{SOUSA: the Swift Optical/Ultraviolet Supernova Archive}
\shortauthors{Brown et al.}

\author{Peter J. Brown}
\affil{George P. and Cynthia Woods Mitchell Institute for Fundamental Physics \& Astronomy, 
Texas A. \& M. University, Department of Physics and Astronomy, 
4242 TAMU, College Station, TX 77843, USA; 
pbrown@physics.tamu.edu}
\and
\author{Alice A. Breeveld}
\affil{Mullard Space Science Laboratory, University College London, Holmbury St. Mary, Dorking Surrey, RH5 6NT, UK} 
\and
\author{Stephen Holland}
\affil{Space Telescope Science Center, 3700 San Martin Dr., Baltimore, MD 21218, USA}
\and
\author{Paul Kuin}
\affil{Mullard Space Science Laboratory, University College London, Holmbury St. Mary, Dorking Surrey, RH5 6NT, UK} 
\and
\author{Tyler Pritchard}
\affil{Department of Astronomy and Astrophysics, 
				The Pennsylvania State University, 
				525 Davey Laboratory, 
				University Park, PA 16802, USA}

\begin{abstract}

The Ultra-Violet Optical Telescope on the Swift spacecraft has observed hundreds of supernovae, covering all major types and most subtypes.  Here we introduce the Swift Optical/Ultraviolet Supernova Archive (SOUSA), which will contain all of the supernova images and photometry.  We describe the observation and reduction procedures and how they impact the final data.  We show photometry from well-observed examples of most supernova classes, whose absolute magnitudes and colors may be used to infer supernova types in the absence of a spectrum.  A full understanding of the variety within classes and a robust photometric separation of the groups requires a larger sample, which will be provided by the final archive.  The data from the existing Swift supernovae are also useful for planning future observations with Swift as well as future UV observatories.

\end{abstract}



\section{Ultraviolet Observations of Supernovae}

Supernova (SN) explosions have been observed in the ultraviolet (UV) since 1972 with the Orbiting Astronomical Observatory (OAO-2; \citealp{Holm_etal_1974}).  In the decades since, UV observations have been made by the International Ultraviolet Explorer (IUE; \citealp{Cappellaro_etal_1995}), the Hubble Space Telescope (HST; e.g. \citealp{Kirshner_etal_1993, Millard_etal_1999, Baron_etal_2000,Foley_etal_2012_11iv}), the Astron Station \citep{Lyubimkov_1990}, the Galaxy Evolution Explorer (GALEX; \citealp{Gal-Yam_etal_2008}), and XMM-Newton's Optical Monitor (OM; \citealp{Immler_etal_2005}).
Atmospheric absorption requires observing from space, so the number of SNe observed in the UV is much lower than in the optical (see \citealp{Panagia_2003, Foley_etal_2008, Brown_2009} for reviews).

\begin{figure}[hb]
\includegraphics[width=8cm]{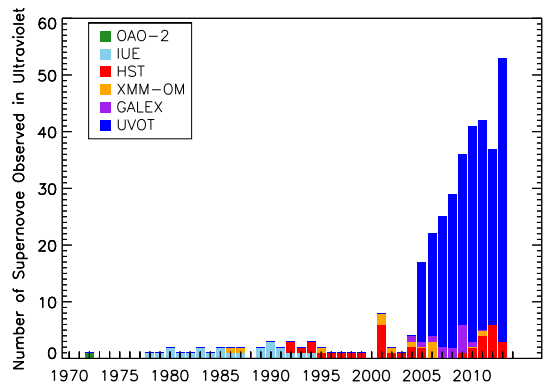}\label{fig_histogram}
\caption{%
Histogram of the number of SNe observed in the UV each year.  Since 2005 the Swift UVOT has observed nearly ten times more SNe than the other missions combined. } 
\end{figure}

\begin{figure*}[h]
\includegraphics[width=16cm]{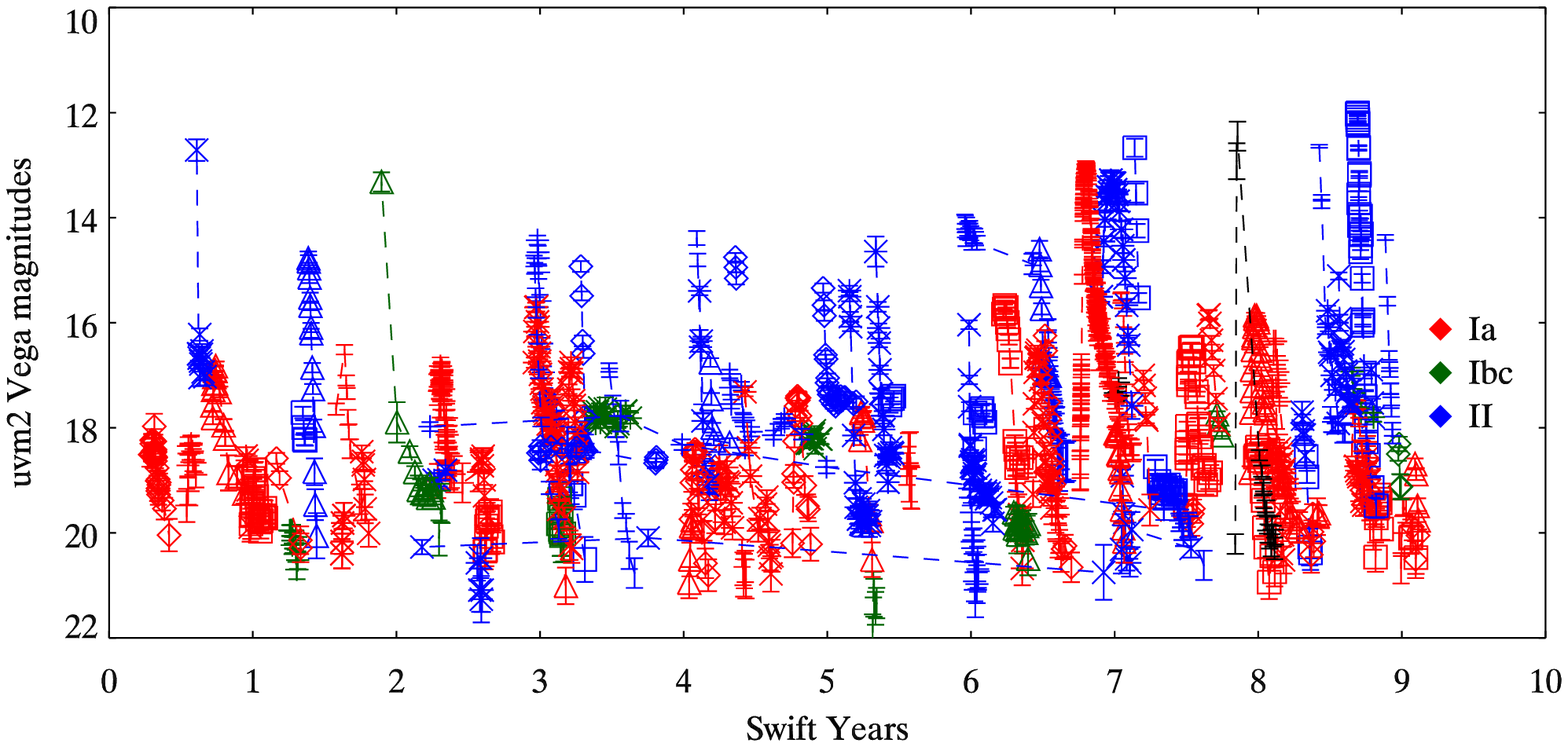}
\includegraphics[width=16cm]{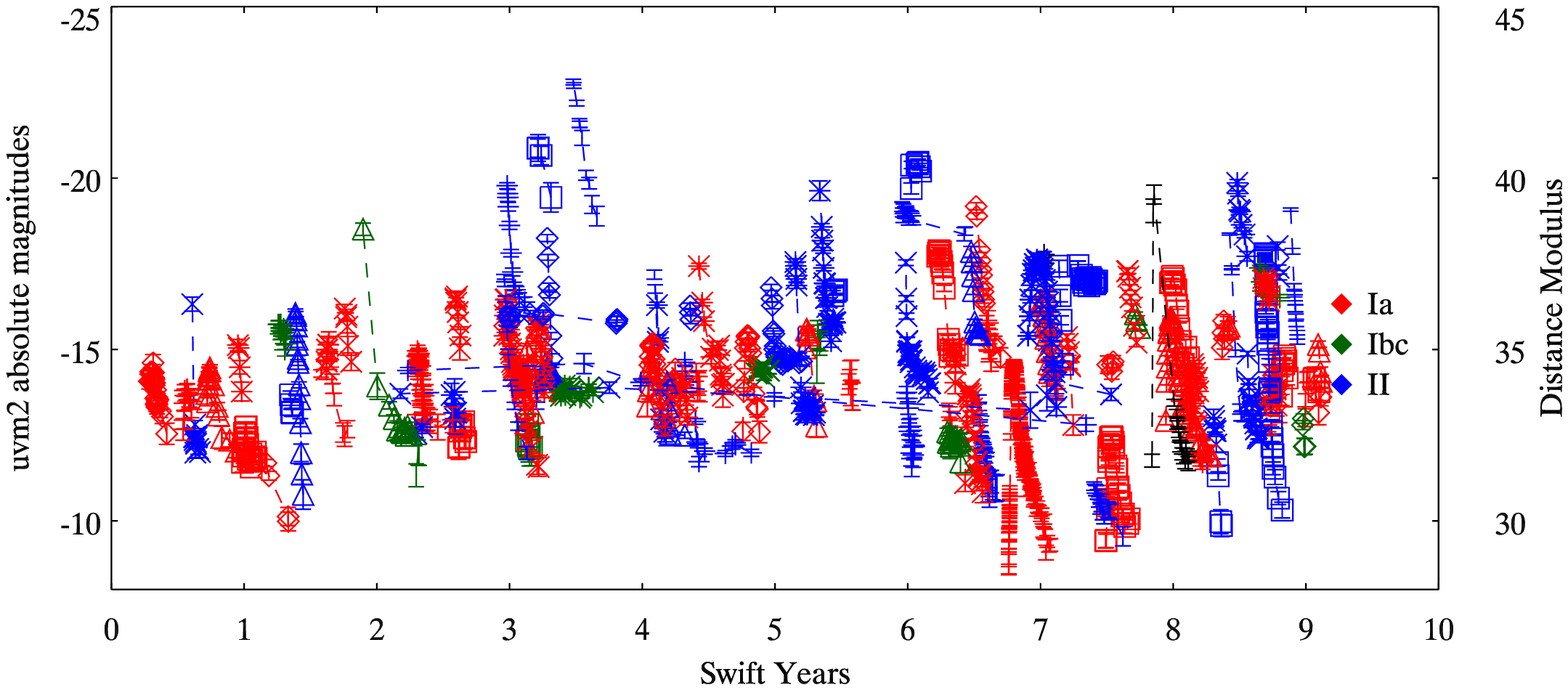}
\caption{%
A timeline of Swift SN observations in the mid-UV ($\sim$2200 \AA) uvm2 filter.  The top plot shows the observed magnitudes.  The bottom panel shows the absolute magnitudes (derived from the host galaxy redshift using H$_0$=72 km/s/Mpc: \citealp{Freedman_etal_2001}).  The y-axis on the right side shows the distance at which that brightness is observable for a limiting magnitude of 20.  
}\label{fig_m2obs} 
\end{figure*}

\clearpage

The Swift Ultraviolet/Optical Telescope (UVOT; \citealp{Gehrels_etal_2004, Roming_etal_2005}) began observing SNe in 2005 \citep{Brown_etal_2005}.  Since then it has observed over 300 SNe.  This dramatic increase in the number of SNe observed in the UV is shown in Figure \ref{fig_histogram}.  The individual SNe are listed on the Swift SN website \footnote{http://swift.gsfc.nasa.gov/docs/swift/sne/swift\_sn.html}.  Many have been published already, including samples in \citet{Brown_etal_2009}, \citet{Milne_etal_2010}, \citet{Brown_etal_2012a}, and \citet{Pritchard_etal_2013}.  Of these, only the latter uses the latest zeropoint calibration and time-dependent flux sensitivity correction of \citet{Breeveld_etal_2011}.  For a better comparison of the growing sample, we are analyzing or reanalyzing all of the UVOT SN data and creating the Swift Optical/Ultraviolet Supernova Archive (SOUSA).  The final archive will include the imaging data as well as the photometry. The Swift/UVOT also has optical and ultraviolet grisms to perform low resolution spectroscopy (\citealp{Roming_etal_2005, Kuin_etal_2009}, Kuin et al. 2014, in preparation).  The grisms have been used for SNe as well \citep{Bufano_etal_2009, Foley_etal_2012_09ig, Bayless_etal_2013, Margutti_etal_2014}, but we focus for now on the photometry.  
 This article is intended to introduce the archive and describe the photometry products that will be released via the Swift SN website.  Data for most SNe previously published by us, notably \citet{Brown_etal_2009} and \citet{Brown_etal_2012b}, are already available and more will be added over the coming months.  In addition, we provide some of the scripts used to reduce the data and parse the output files.  In Section \ref{section_obs} we describe the Swift observations and in Section \ref{section_data} we detail the photometric reduction.  In Section \ref{section_results} we use some of the photometry to show how the different SN classes differ in UV colors and absolute magnitudes and how it can be used to plan future observations with Swift and future UV observatories.

\begin{figure}[t]
\includegraphics[width=8cm]{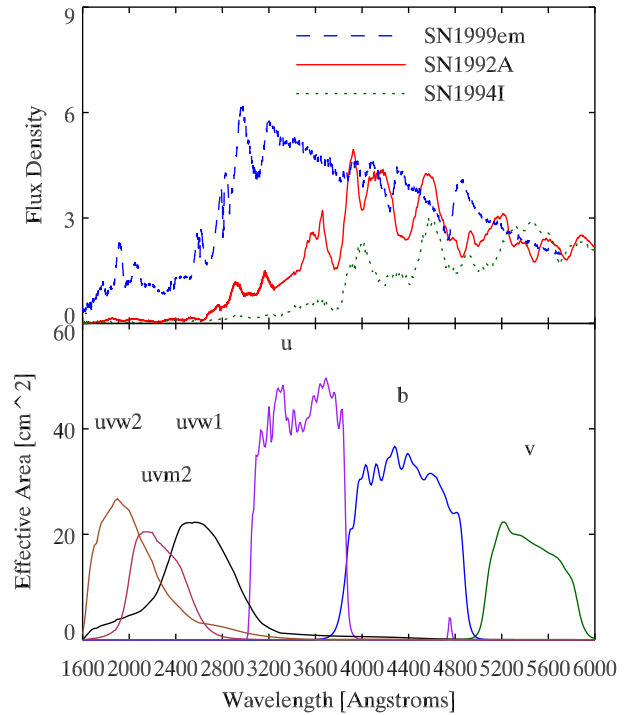}\label{fig_filters}
\caption{%
Top panel: the effective areas of the UVOT filter curves.  Bottom panel: UV/optical spectra of SNe from HST, normalized to have the same optical flux to show the diversity in the UV flux levels.\label{fit_histogram} } 
\end{figure}

\section{Swift UVOT Observations\label{section_obs}}

The Swift spacecraft \citep{Gehrels_etal_2004} was designed for the detection and rapid observation of gamma ray bursts (GRBs).  
It has a special capability whereby a target position can be uploaded to the spacecraft for immediate observation whenever viewable, superseding the previously planned targets.  This allows newly discovered SNe to be observed within hours of discovery.  The data is regularly sent down from the spacecraft and usually available from the Swift website\footnote{http://swift.gsfc.nasa.gov/cgi-bin/sdc/ql?} several hours later.  This allows for rapid feedback on the UV brightness of a new target to inform the planning of future observations, which is usually created just one or two days in advance.  Swift observes several targets during its 90-minute orbit, so the overhead on individual targets is low compared to other space observatories.  This allows many relatively short observations to be scheduled to obtain better time coverage than usually possible in the UV.  These unique features make Swift an excellent observatory for transients such as SNe.

With a few notable exceptions (the GRB-SN~2006aj: \citealp{Campana_etal_2006} and the shock breakout of SN~2008D: \citealp{Soderberg_etal_2008}) Swift does not discover SNe.  SNe discovered elsewhere are proposed as targets of opportunity (ToOs) and, if approved, subsequently observed.  Because most are proposed one by one (with the exception of some guest investigator programs), there is not a uniform selection criteria.  We have not tried to obtain an unbiased sample of all SNe but to obtain observations of SNe across all types and host galaxy environments as much as possible.  Because of the UV faintness of many SN types and the relatively small aperture of UVOT, most are very nearby SNe, with a redshift of z less than 0.02 (but we will discuss limits on this later).  SNe without  significant extinction or galaxy contamination are usually preferred.  SNe have typically been observed with a two-day cadence during the early phases and spreading out as the SN ages and changes less with time. After the SN has faded, an observation of the host galaxy is requested as a low priority target that can be filled in as the schedule allows. The excellent temporal coverage is reflected in Figure \ref{fig_m2obs} which shows preliminary data on most of the SNe observed over Swift's lifetime.

For clarity we will now define a few of the terms we use in the observing and data analysis.  An ``observation'' typically refers to one or more exposures scheduled as a set in the pre-planned science timeline (PPST) or executed by the spacecraft as the result of an uploaded command to immediately observe an ``automated target'' (AT).  An observation may include multiple exposures in different filters and may span several orbits.  As Swift is in a low Earth orbit, locations in the sky are not observable for large continuous chunks of time (typically not more than 40 minutes).  Long observations are broken up into ``snapshots'', continuous viewing periods during which exposures are taken in a predetermined sequence of UVOT filters.

The Swift UVOT observations usually utilize the six main broadband filters (see \citealp{Roming_etal_2005} and \citealp{Poole_etal_2008} for details).  The UVOT filters are compared to SN spectra in Figure \ref{fig_filters}.  The white (clear) filter is not used due to its broad passband which is hard to flux calibrate for objects of different or varying spectral shape.
The UVOT filter mode determines which filters will be used and the exposure times in each.  For a scaled mode, the exposure times in each filter are calculated based on the exposure time ratios given by the mode and the calculated length of the snapshot.  For planned targets the snapshot length is calculated by the spacecraft based on the planned time on target.  For ATs not in the planned timeline the snapshot length is calculated as the time until an observing constraint is reached.  Exposures in some filters may not be observed if the full, planned snapshot is not observed due to a higher merit AT becoming visible during the snapshot.  If a higher merit AT causes the snapshot to begin late, the exposure times in the filters will be calculated based on the time remaining and all requested filters will be observed (albeit for a shorter than planned amount of time).  ATs can also be superseded by higher merit targets in the PPST, resulting in a truncated snapshot which is shorter than what the spacecraft would calculate based on the observing constraints.  To get all filters for prompt SN observations ``unscaled'' modes can be used. In these modes the filters are observed for a set amount of time so that all filters can be completed within the snapshot (whose length can be determined beforehand).

Here we highlight the recommended modes for SN photometry: the preferred PPST mode for red objects (like most SNe) is the scaled mode 0x223f, which has the six UVOT filters (all of the broadband filters except for white) with the following approximate time fractions (uvw1,u,b,uvw2,v,uvm2) (17,8,8,25,8,33).  The preferred AT mode is the unscaled mode 0x0270 with times of (uvw1,u,b,uvw2,v,uvm2) (160 s, 80 s, 80 s, 320 s, 80 s, 280 s) for a 1000 s snapshot.  If the snapshot is longer than 1000 s the remaining time is spent in the uvm2 filter, usually valuable for UV-faint targets.  

UVOT uses a photon-counting detector.  As such, the count rates from sources brighter than $\sim$13 mag cannot be accurately measured in the normal modes and photometric procedures.  Special ``hardware-window'' modes read out a smaller portion of the detector.  The faster readout means higher temporal resolution so that the count rate can be determined for sources as bright as $\sim$12th magnitude.  Recently, a method to use the readout streak of bright point sources was developed \citep{Page_etal_2013} which can provide photometry up to $\sim$10th magnitude (with an uncertainty of about 0.1 mag) provided the readout streak is well exposed.

\section{Data Analysis\label{section_data}}

\subsection{Image Processing}

Processed images are downloaded from the Swift archive\footnote{http://heasarc.gsfc.nasa.gov/cgi-bin/W3Browse/swift.pl}.  The archive is searched using the SN position so that all images of the field are obtained regardless of the target identification number (TID).  Sometimes multiple TID numbers are used to differentiate different programs or observations made of the galaxy rather than targeting the SN.  We use the sky images which are shifted and rotated into the World Coordinate System.  Each  fits file (suffix .img) contains all exposures in a given filter for that observation corresponding to a unique observation identification number (OBSID).\footnote{See http://archive.stsci.edu/swiftuvot/file\_formats.html for a more detailed description.}  
For bright SNe like 2011fe \citep{Brown_etal_2012b}, individual exposures are used.  Otherwise, all full-field exposures within a single OBSID are coadded into a single image for that epoch.  Exposures using different frame rates are not coadded because the coincidence losses (and corrections) are different.  Images are examined so that individual exposures that show image artifacts (such as streaking stars due to the spacecraft moving during the exposure) can be excluded.  We do not require the aspect correction to have been successfully performed but manually correct or exclude any images that are offset from the rest.  A new fits file is created for each filter with extensions including the summed image from each epoch.  A separate fits file is created for the images designated as templates which were observed before the SN exploded or after it had significantly faded.

\subsection{Photometry}

The photometric reduction follows the same basic outline as \citet{Brown_etal_2009}.  
\texttt{HEASOFT} (currently version 6.13 corresponding to Swift release 4.0) is used to perform the photometry using \texttt{uvotsource}.  It is called by \texttt{uvotmaghist} which operates on a list of images or a single fits file with multiple image extensions.  It creates as output a fits table of the extracted and calculated values.  The calibration database (CALDB) version released 2013-01-18 is used, which includes the zeropoints in the UVOT Vega and AB systems \citep{Breeveld_etal_2011}.  The default photometry in SOUSA is on the UVOT-Vega system, while conversion to the AB system is straightforward using the zeropoint differences in \citep{Breeveld_etal_2011}.
Counts in the source region are measured using 3\arcsec~and 5\arcsec~apertures.  The coincidence loss correction for the source is determined using the 5\arcsec aperture.  The coincidence loss is computed separately for the background.  The source counts are obtained by subtracting the coincidence-loss corrected background (scaled for the size of the aperture) from the corrected total counts in the source aperture.   The count rates are also corrected for the time-dependent loss in sensitivity \citep{Breeveld_etal_2011} which amounts to 1\%~per year in all filters except v which is now corrected by 1.5\% per year\footnote{http://heasarc.gsfc.nasa.gov/docs/heasarc/caldb/swift/docs/uvot/uvotcaldb\_throughput\_02b.pdf}. 
Necessary corrections to the exposure time include subtracting the time during which the frames are being downloaded and rare anomalies.\footnote{http://swift.gsfc.nasa.gov/analysis/uvot\_digest/timing.html}

If individual images are used, a correction is made for differences in the large scale sensitivity \citep{Breeveld_etal_2010}.  When photometry is done on coadded images (where the source does not correspond to a unique detector position) the correction is not done, and a systematic uncertainty of 2.3\% of the count rate is added in quadrature to the photometric error \citep{Poole_etal_2008}.

The above steps are done for each of the SN images as well as the summed template image, giving corrected count rates in the 3\arcsec~and 5\arcsec~apertures.  These are taken from the fits file output of \texttt{uvotmaghist} and then corrected for the galaxy using our own scripts as follows. The count rates from the galaxy template in the appropriate aperture are subtracted from the count rates in the SN images.  This is done before the aperture correction since the SN is a point source and the galaxy background is likely not.  For the 3\arcsec~aperture the galaxy count rate is subtracted and then the aperture correction from \texttt{uvotmaghist} (calculated using the average UVOT PSF in the CALDB) is applied.

In choosing the aperture size there is a trade-off between maximizing the signal to noise ratio for faint objects \citep{Li_etal_2006_UVOT,Poole_etal_2008} and the uncertainties of the correction to the full photometric aperture.  We calculate photometry using 3\arcsec ~and 5\arcsec~apertures.  A 1.5\% uncertainty on the flux is added in quadrature to the error when using the 3\arcsec~aperture to account for variations in the point spread function \citep{Breeveld_etal_2010}.  For each photometric point we choose the aperture with the smallest magnitude error.

\begin{figure}
\hfill
\subfigure[Title A]{\includegraphics[angle=0,width=8cm]{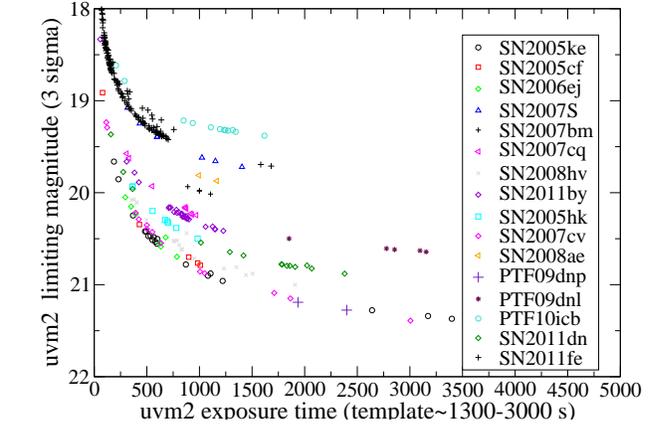}}
\hfill
\subfigure[Title B]{\includegraphics[angle=0,width=8cm]{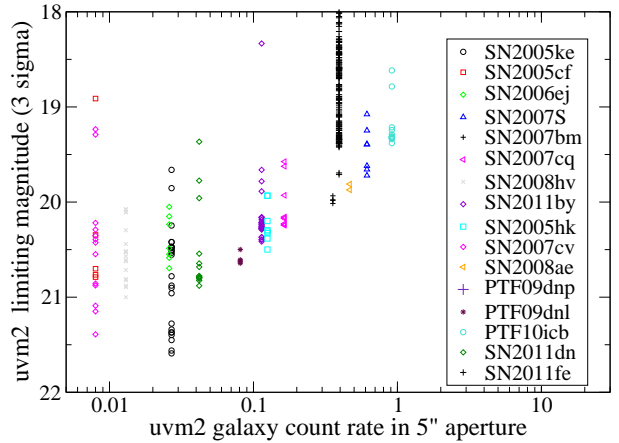}}
\hfill
\caption{Upper limits for a sample of SNe as a function of the exposure time (top panel) and the count rate of the underlying galaxy (bottom panel).  The limits level out beyond 1000 s and are dominated by the brightness of the galaxy. \label{fig_upperlimits}}
\end{figure}

Upper limits are calculated from the number of counts required to achieve a signal to noise (S/N) of three when accounting for the statistical error on the source counts and the errors on the background and galaxy counts.  Using a Poisson error  rather than the binomial error appropriate for photon counters \citep{Kuin_Rosen_2008} makes this analytically possible and is a good approximation in the low count regime.  The count limit is corrected for the aperture size, large scale sensitivity, and time dependent sensitivity and converted to a magnitude.  This limiting magnitude is given in the data table.  Magnitudes falling below this S/N=3 limit are removed from the data table.  The count rates and errors are still given for those epochs, as they are more useful for constraining models than the upper limits. The upper limits are a function of the exposure time, the background count rates, and the galaxy count rates.  As shown in Figure \ref{fig_upperlimits}, above an exposure time of 1000 s the galaxy count rate dominates the upper limit.  Because the large scale sensitivity and PSF uncertainties (which scale with the count rate) are propagated into the error, the underlying galaxy count rate imposes a floor on how faint a source could be significantly detected.  Image subtraction techniques may alleviate some of these problems but would also have to deal with coincidence loss issues which may be significant for the extended galaxy light \citep{Breeveld_etal_2010} even if the SN itself is faint.  For the standard UV-weighted UVOT mode 0x223f, 1/3 of the time is the uvm2 filter.  One can estimate the time needed in uvm2 and multiply by three to estimate the exposure time needed for all six filters.
\clearpage

Some sources are bright enough to saturate UVOT's photon-counting detector, meaning that nearly every frame is recording a count such that the true number of incident photons cannot be determined.  For each epoch we set the limiting magnitude on the bright side (given the background and galaxy counts) corresponding to a measured count rate of 0.98 counts per frame and report that in the data table.  For count rates above this limit, both the magnitude and the count rate are excluded from the data table.  The count rate errors for saturated sources are not useful as constraints as the upper bound is infinite. 

The accuracy of the photometry has been checked using a variety of SN and non-SN sources \citep{Poole_etal_2008}.  
We restrict comparisons to ground-based data to B and V due to the shorter wavelength response of the Swift u filter compared to ground-based Johnson U and Sloan u.  Where differences have been found, it is usually due to a nearby star (as in the case of SN~2005am) or a high count rate from the underlying galaxy which causes the coincidence loss correction to be underestimated (SNe 2006dd, 2006mr, 2011iv, others).  Based on comparisons, we exclude data where the underlying galaxy is measured to be brighter than 8 counts/s with a caution that the photometry might be off by 0.05 mag between 6-8 counts/s.  The coincidence loss from a flat, extended source was studied in \citet{Breeveld_etal_2010} but is not known for the case of a structured background like a galaxy.  In the absence of such issues, we find the UVOT b,v photometry generally agree within 0.05 mags of published ground-based B,V photometry even without accounting for small differences in the filter shapes.

\clearpage
\subsubsection{SN2011fe}

\begin{figure}[ht]
\includegraphics[width=8.0cm]{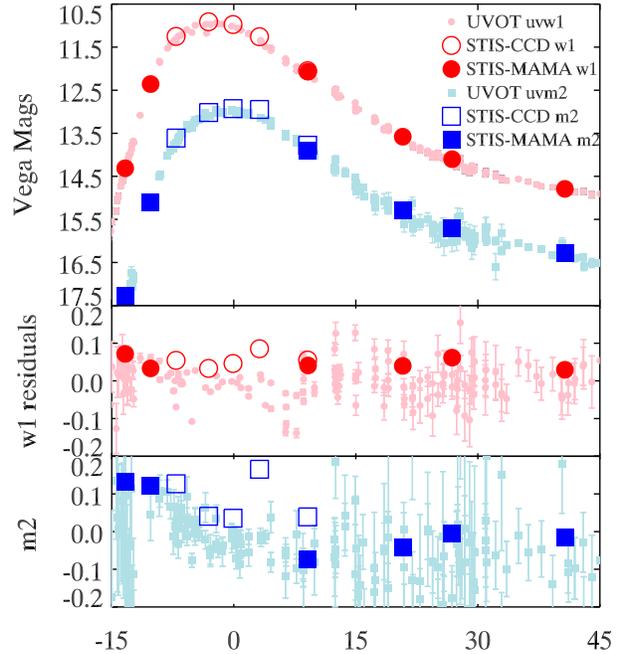}
\caption{%
UV light curves of SN~2011fe with UVOT photometry in the uvw1 and uvm2 bands and spectrophotometry from HST/STIS CCD and MAMA spectra.  
The bottom panels show the residuals from a polynomial model \citep{Brown_etal_2012b} to flatten the curves and allow a visual comparison between the UVOT photometry and the HST spectrophotometry. It is not the y-value that is important but the consistency between the UVOT and HST points. } \label{fig_11fe}
\end{figure}

The very nearby SN~2011fe is the best-studied SN Ia at all wavelengths.  In the UV, it was observed within two days of explosion \citep{Nugent_etal_2011} with a well-sampled light curve from UVOT \citep{Brown_etal_2012b}.  \citet{Pereira_etal_2013} found a 0.2 mag discrepancy in all 6 UVOT bands compared to their HST and Supernova Factory spectrophotometry.  Most of this difference is the result of the time-dependent sensitivity not being uniformly applied to the photometry in \citet{Brown_etal_2012b}.  To further check the consistency, we have downloaded the HST spectra of SN~2011fe from the HST/MAST archive \citet{Mazzali_etal_2013} and performed our own spectrophotometry in the UVOT system. The spectra used by \citealp{Pereira_etal_2013} were interpolated in time to their optical spectra for the creation of a bolometric light curve.  
After correction for the UVOT time-dependent sensitivity, the HST UV spectrophotometry is generally within the scatter of the UVOT photometry.  This is shown in Figure \ref{fig_11fe}.  In the uvm2 filter the HST-STIS/CCD spectrophotometry is still 0.1 mag brighter than the UVOT photometry and the HST-STIS/MAMA spectrophotometry for the one epoch with observations in both.  This is likely due to scattered light in the STIS/CCD.\footnote{See Figure 6 in http://archive.stsci.edu/pub/hlsp/stisngsl/aaareadme.pdf}

\clearpage

\begin{figure*}[ht]
\includegraphics[width=16cm]{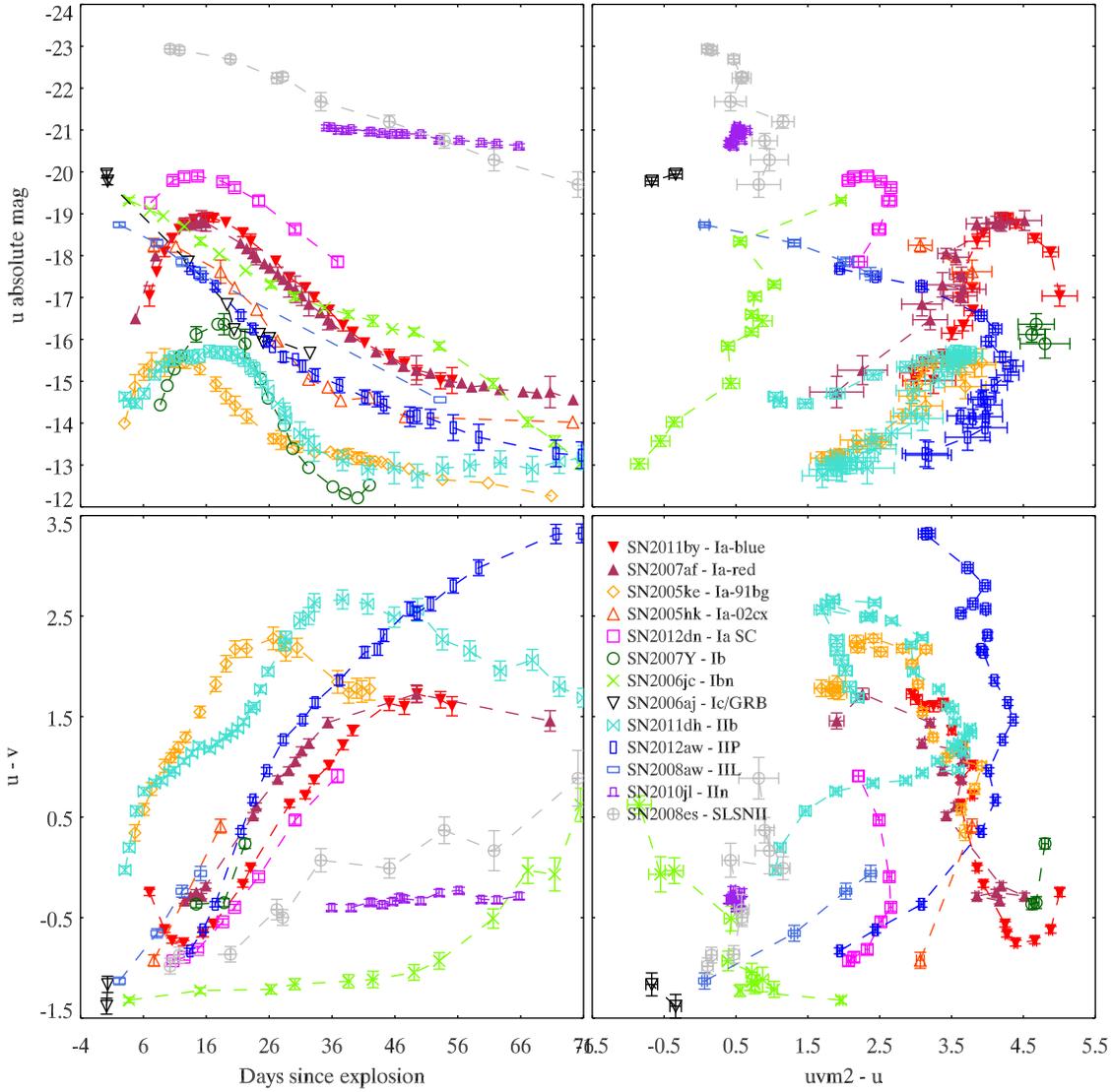}
\caption{%
Top left: u-band absolute magnitudes versus estimated time since explosion.  Top Right: u-band absolute magnitudes versus uvm2-u colors.
Bottom left: u-v colors versus time.  Bottom right: u-v colors versus uvm2-u colors.} \label{fig_colors}
\end{figure*}

\section{Sample Application\label{section_results}}

The well-sampled, multi-filter Swift SN data is excellent for studying the UV and optical evolution of individual SNe and comparing across different classes.  The sample size is also large enough to compare objects within the same class or subclass \citep{Milne_etal_2013, Pritchard_etal_2013}.   The Swift sample also includes some rare objects and subclasses that can be compared to the others.  One potentially fruitful application is using the UV/optical photometry to distinguish the SN class or even subclass without requiring spectroscopy.  

\citet{Cappellaro_etal_1995} and \citet{Panagia_2003} used IUE spectra/spectrophotometry to show the UV color differences between SNe I and II.
In \citet{Brown_etal_2009} we used the Swift/UVOT data to add the temporal dimension to show that SNe IIP are only bluer than SNe I at early times.  SNe IIP become redder with time, becoming indistinguishable in color beginning about two weeks after explosion.
The situation has become more complicated with the increase in subclasses observed by Swift/UVOT. \citet{Brown_etal_2009} did not include SNe IIL, IIn, or IIb, the recently identified classes of super-Chandrasekhar mass SNe Ia or superluminous SNe (SLSNe; \citealp{Gal-Yam_2012}).  The intrinsic dispersion of colors within a class of SNe is also better understood with a larger sample and may lead to the identification of differences or subclasses within a class \citep{Milne_etal_2013}.

In Figure \ref{fig_colors}, we revisit some of the color-color and color evolution plots from \citet{Brown_etal_2009} using well-observed, local SNe of most classes and subclasses.  Figure \ref{fig_colors} shows the time evolution of the u-band absolute magnitudes and u-v colors, an absolute u band magnitude versus uvm2-u color plot, and a u-v versus uvm2-u color-color plot.  Given a SN type and current optical magnitude, one can estimate by the color of similar objects the current and future UV brightness and thus the observability by Swift or a future UV mission.  Adding absolute magnitude as a dimension breaks some degeneracies of color and extinction. Some regions of some plots are more congested than others, but rarely are the same degeneracies present in all the plots. 
If the colors of an object match more than one type, usually the addition of multiple epochs will allow the object to be uniquely typed.   For real-time adaptation of observing plans, Figure \ref{fig_colors} or something similar may be sufficient (see also \citealp{Gal-Yam_etal_2004} for an example of optical phototyping).  When the classification needs are more rigorous (i.e. for identification of cosmologically useful SNe Ia or differentiating core-collapse and thermonuclear SNe for rate measurements), a larger sample needs to be utilized to include the dispersion within classes and a statistical treatment of the likelihoods.  The full version of SOUSA will be an excellent data set of rest-frame UV photometry against which SNe can be compared.  Photometric classification of SNe will be critical for large surveys such as LSST which will find many more SNe than can be followed up spectroscopically, and many of these will be observed in the rest-frame UV.

From the absolute magnitudes one can also determine the distance out to which one can follow desired phases of different SN types with Swift/UVOT or future UV observatories (see also Figure \ref{fig_m2obs} for the mid-UV apparent and absolute magnitudes).  For most SN types the limiting distance is farther than the z=0.02 commonly observed in the past with Swift.  SNe Ia are now being targeted between z=0.02-0.035 in the nearby Hubble flow to improve their distance and absolute peak magnitude measurements.  Young SNe II could be observed even farther. 
Several UV-bright SLSNe can be seen rising above the rest of the SNe in the bottom panel of Figure \ref{fig_m2obs}. These include the hydrogen-rich SNe~2008es \citep{Gezari_etal_2009,Miller_etal_2009} and 2008am \citet{Chatzopoulos_etal_2011} and several hydrogen-poor SLSNe (PTF09atu, PTF09cnd, and PTF09cwl; \citealp{Quimby_etal_2011}) which were observed at redshifts z$\sim0.2$.  The extreme brightness of SN~2008es suggests similar objects could be detectable by Swift/UVOT out to redshifts of z$\sim0.5$.  This would provide rest-frame observations near Lyman alpha for comparison with the extremely high-redshift SNe that are now being found in deep optical searches \citep{Cooke_etal_2012}.  The current and future Swift SN observations will provide a legacy for future observations of SNe near and far in the rest-frame UV.

\acknowledgments

We wish to thank the many members of the Swift team who respond to and schedule our many SN requests, sometimes at very inconvenient times.  We are also grateful to the many different groups who discover SNe and announce them to the world so we can follow them up in the UV.
SOUSA is supported by NASA's Astrophysics Data Analysis Program through grant NNX13AF35G.

\nocite{*}
\bibliographystyle{spr-mp-nameyear-cnd}
\bibliography{biblio-u1}

{}

\end{document}